\documentclass[sigconf]{acmart}
\AtBeginDocument{%
  \providecommand\BibTeX{{%
    \normalfont B\kern-0.5em{\scshape i\kern-0.25em b}\kern-0.8em\TeX}}}

\copyrightyear{2023}
\acmYear{2023}
\setcopyright{acmlicensed}
\acmConference[CIKM '23] {Proceedings of the 32nd ACM International Conference on Information and Knowledge Management}{October 21--25, 2023}{Birmingham, United Kingdom.}
\acmBooktitle{Proceedings of the 32nd ACM International Conference on Information and Knowledge Management (CIKM '23), October 21--25, 2023, Birmingham, United Kingdom}
\acmPrice{15.00}
\acmISBN{979-8-4007-0124-5/23/10}
\acmDOI{10.1145/3583780.3615142}
\usepackage[ruled,vlined]{algorithm2e}
\usepackage{multirow}
\usepackage{booktabs,caption}
\usepackage[flushleft]{threeparttable}
\usepackage{latexsym}
\usepackage{amsmath}

\begin{document}

\title{Knowledge-Enhanced Multi-Label Few-Shot Product Attribute-Value Extraction}

\author{Jiaying Gong}
\affiliation{%
  \institution{Virginia Polytechnic Institute and State University}
  \city{Blacksburg}
  \country{U.S.}}
\email{gjiaying@vt.edu}

\author{Wei-Te Chen}
\affiliation{%
  \institution{Rakuten Institute of Technology, \\ Rakuten Group Inc.\\}
  \city{Boston}
  \country{U.S.}}
\email{weite.chen@rakuten.com}

\author{Hoda Eldardiry}
\affiliation{%
  \institution{Virginia Polytechnic Institute and State University}
  \city{Blacksburg}
  \country{U.S.}}
\email{hdardiry@vt.edu}

\renewcommand{\shortauthors}{Jiaying Gong, Wei-Te Chen \& Hoda Eldardiry}

\begin{abstract}
Existing attribute-value extraction (AVE) models require large quantities of labeled data for training.
However, new products with new attribute-value pairs enter the market every day in real-world e-Commerce.
Thus, we formulate AVE in multi-label few-shot learning (FSL), aiming to extract unseen attribute value pairs based on a small number of training examples.
We propose a Knowledge-Enhanced Attentive Framework (KEAF) based on prototypical networks, leveraging the generated label description and category information to learn more discriminative prototypes.
Besides, KEAF integrates with hybrid attention to reduce noise and capture more informative semantics for each class by calculating the label-relevant and query-related weights.
To achieve multi-label inference, KEAF further learns a dynamic threshold by integrating the semantic information from both the support set and the query set.
Extensive experiments with ablation studies conducted on two datasets demonstrate that KEAF outperforms other SOTA models for information extraction in FSL.
Code is: \href{https://github.com/gjiaying/KEAF}{https://github.com/gjiaying/KEAF}
\end{abstract}




\maketitle
\section{Introduction}
Product attribute value pairs are important for e-Commerce because platforms make product recommendations for customers based on the key attribute-value pairs and customers use attributes to compare products and make purchases.
Existing studies on AVE based on neural networks view AVE as sequence labeling~\cite{yan-etal-2021-adatag, jain-etal-2021-learning}, question-answering~\cite{wang2020learning, shinzato2022simple} or multi-modal fusion problems~\cite{zhu-etal-2020-multimodal, lin2021pam}.
These supervised-learning models are well trained to classify attribute-value pairs when large quantities of labeled data are available for training.
Even the most current open mining model needs a few attribute-value seeds and iterative training for weak supervision~\cite{10.1145/3485447.3512035}.
However, new products with new attribute-value pairs enter the market every day in real-world e-Commerce platforms.
It is difficult, time-consuming and costly to manually label large quantities of new products profiles for training.
Besides, with the appearance of new attribute-value pairs, the class distribution becomes long-tailed, where a subset of the labels have many samples, while majority of the labels have only a few samples.

\begin{figure}[htp] 
  \vspace{-0.125in}
 \center{\includegraphics[height=5.1cm,width=8.2cm]{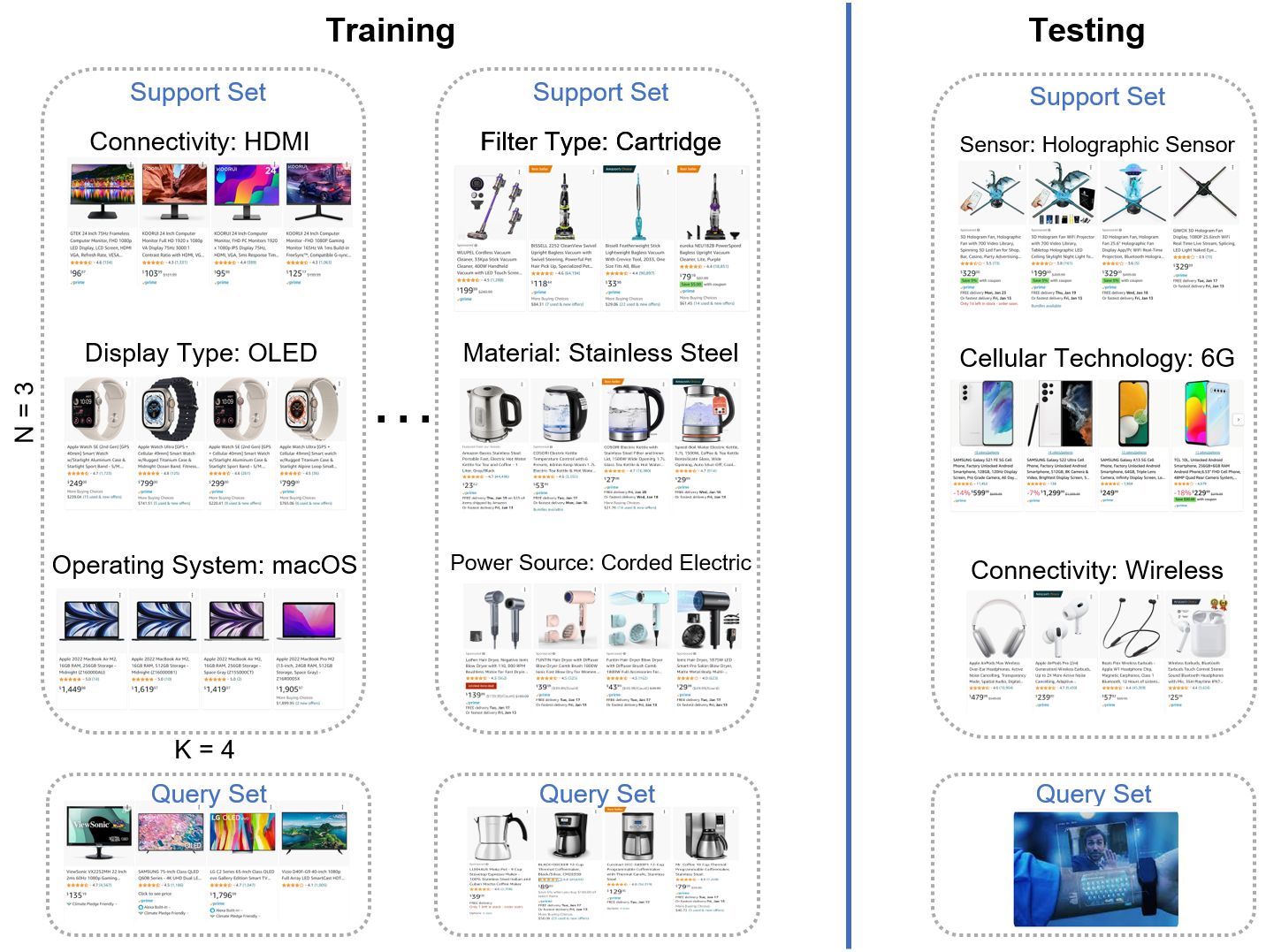}}
 \caption{\label{fig:example} An example of multi-label few-shot product attribute-value extraction task.}
 \vspace{-0.125in}
 \end{figure}
 
We formalize AVE as a multi-label FSL problem, aiming to extract structured product information from unstructured profiles with limited training data.
We take the common head labels data for training and the limited tail labels data for testing, and there is no overlap of classes between training set and testing set shown in Figure~\ref{fig:example}.
Recent methods on multi-label FSL have made great progress in CV~\cite{alfassy2019laso, simon2022meta} and NLP~\cite{hou2021few, liu2022label, zhao2022label}. Among these methods, prototypical network~\cite{snell2017prototypical} has been proved to be powerful and potential.
However, different from AVE in e-Commerce, these models (1) explore only label tags as auxiliary information, (2) still have noise when learning prototypes, and (3) require further data or additional models to learn the threshold for label numbers prediction.

To address the above challenges, we propose a Knowledge -Enhanced Attentive Framework (KEAF) for product AVE. 
The main contributions of KEAF consist of three parts.
(1) To the best of our knowledge, we are the first to formulate AVE as a multi-label FSL task to tackle the problem of limited training data for long-tailed datasets.
Unlike open mining models, KEAF does not require the attribute-value pairs exactly appear in the product profiles.
(2) By leveraging both the label description generated by a generator and the category information as the auxiliary information to obtain more discriminative prototypes, KEAF can not only avoid the issue that different attribute-value pairs share the identical prototype for 1-shot learning, but also alleviate ambiguity by obtaining both label and category relevant information.
The hybrid attention mechanism also helps reduce the noise and capture more informative semantics from the support set by calculating both the label-relevant and query-related weights.
(3) To achieve multi-label inference, a dynamic threshold is learned during the training stage by integrating the semantic information from support and query sets. 
The adaptive threshold does not require additional training data or based on additional models.
Extensive experimental results on two datasets show that our proposed model KEAF significantly outperforms other existing information extraction models for AVE.


\section{Related Works}
Early works on AVE use a domain-specific dictionary and rule-based methods to identify attribute value pairs~\cite{10.5555/2145432.2145598, shinzato-sekine-2013-unsupervised, 34460, 10.1145/1147234.1147241}.
Then, sequence labeling models~\cite{8731553, 10.1145/3219819.3219839, yan-etal-2021-adatag, jain-etal-2021-learning}, question answering-based models~\cite{xu2019scaling, wang2020learning, shinzato2022simple}, multi-modal models~\cite{zhu-etal-2020-multimodal, lin2021pam, wang2022smartave}, extreme multi-label models~\cite{chen2022extreme, 10020304} and open mining models~\cite{10.1145/3485447.3512035} are trained for AVE.
However, these approaches require large quantities of labeled data for training or iterative training for weak supervision.
Most works of FSL focus on single-label classification ~\cite{koch2015siamese, baldini-soares-etal-2019-matching, han-etal-2021-exploring, ijcai2022p407, liu-etal-2022-simple}. 
However, one product may have multiple attribute value pairs for AVE task.
Early works on multi-label FSL depends on a known structure of the label spaces~\cite{rios2018few} and label set operations~\cite{alfassy2019laso}
Then, prototypical networks~\cite{NIPS2017_cb8da676} are revised for multi-label cases by learning a shared embedding space~\cite{yang2019prototypical}, grouping samples multiple times~\cite{simon2022meta}, and learning local features with different labels~\cite{yan2022inferring}.
Attention mechanisms~\cite{DBLP:conf/acl/HuZGXGGCS20} and label information~\cite{hou2021few, liu2022label, zhao2022label} are considered to differentiate prototypes.
Different from these approaches, we leverage both label and category for product AVE in e-Commerce.

\section{Methodology}
\subsection{Problem Definition}~\label{sec:definition}
Given a set of training classes $Y_{train}$ and testing classes $Y_{test}$, where $Y_{train}\cap Y_{test}=\varnothing $.
The model is trained with numerous samples from $Y_{train}$, and it can quickly adapt to $Y_{test}$ with few labeled data.
Each training episode involves a support set $S=\begin{Bmatrix}(x_{i}, y_{i})\end{Bmatrix}_{i=1}^{N_{s}}$ and a query set $Q=\begin{Bmatrix}(x_{i}, y_{i})\end{Bmatrix}_{i=1}^{N_{q}}$, where $S$ usually includes $K$ samples (K-shot) for each of $N$ labels (N-way).
In contrast to the single label N-way-K-shot setting~\cite{10.1145/3386252}, multi-label FSL allows that each single sample can have multiple labels simultaneously.
There are $N$ total classes, and each class has at least $K$ samples (with at least one label appearing less than $K$ times if any samples are removed) because we can not guarantee each label appears exactly $K$ times while each sample has multiple labels.
The input data for each product $x$ is a tuple $<t, d, l, c>$, where $t$ is the product title, $d$ is the product description, $l$ is the label description, and $c$ is the product category.
The input label is a vector $y=\begin{Bmatrix}y_{1}, y_{2}, \dotsc, y_{N}\end{Bmatrix}$, where $y \in \begin{Bmatrix}0, 1\end{Bmatrix}$, indicating whether the product has the label or not, and $N$ is the total number of classes.
The outputs are attribute-value pairs.

\subsection{Multi-label Few-Shot Data Sampling}~\label{sec:sampling} 
Multi-label few-shot data sampling includes data splitting, data balancing and data sampling.
We first reconstruct the dataset by splitting data based on upper thresholds $t_u$ and lower thresholds $t_l$, learned from the frequency of class labels to guarantee that $Y_{train}\cap Y_{test}=\varnothing $.
We filter the dataset by discarding the samples with the label count below $t_l$ or above $t_u$, updating the label dictionary and its corresponding samples.
To guarantee that the shot $K_{S}+K_{Q} >= 10$ for FSL, the filtering process is done iterately until class number $N$ is fixed.
Then, we balance the dataset by randomly dropping single-label data to achieve a similar size with multiple-label data. 
To approximately conduct N-way-K-shot learning, we follow~\cite{hou2021few} to construct query and support sets for each episode.
Details of multi-label few-shot data sampling are shown in Algorithm~\ref{alg:data_sampling}.

\begin{algorithm}[!htbp]
\small
\SetKwInOut{KIN}{Input}
\SetKwInOut{KOUT}{Output}
\caption{Multi-label Few-shot Data Sampling}
\label{alg:data_sampling} 
\KIN{Dataset $X$, label set $Y$, support shot $K_S$, query shot $K_Q$, upper threshold $t_u$ and lower threshold $t_l$}
\KOUT{Support set $S$, query set $Q$, query label set $Q_L$}
Initialize $S = \left \{  \right \}$, $Q = \left \{  \right \}$, $Q_L=\left [  \right ]$ and Dict $\begin{Bmatrix}label:count\end{Bmatrix}$ \\
\While {len($Y$) is not fixed}
{
    \uIf {Count($Y_{X_{i,j}}$) $>$ $t_u$ or Count($Y_{X_{i,j}}$) $<=$ $t_l$}
    {
    $X$.remove($X_{i,j}$)
    }
    
    $Y$.update($X$)\\
    \uIf {get\_class($X_{i,j}$) $\notin$ $Y$}
    {
    $X$.remove($X_{i,j}$)
    }
}
data\_balancing($X$)\\
\For{$i$ in Enumerate($Y$)}
{
    indices = $Random(Y_i, K_S+K_Q)$;  count = 0\\
   
    \For{$j$ in indices}
    {
        \eIf {count $<$ $K_Q$}
        {
            $Q$.update($X_{i,j}$); $Q_L$.append($Y_{i,j}$)
            
        }
        {
        \uIf {any Dict$\begin{bmatrix}Y_{i,j}\end{bmatrix} < K_S$}
        {
         $S$.update($X_{i,j}$); Update Dict
        }

        }
        count++
    }
    
}
return $S$, $Q$, $Q_L$
\end{algorithm}

\begin{figure*}[htp] 
 \center{\includegraphics[height=5.2cm,width=15cm]{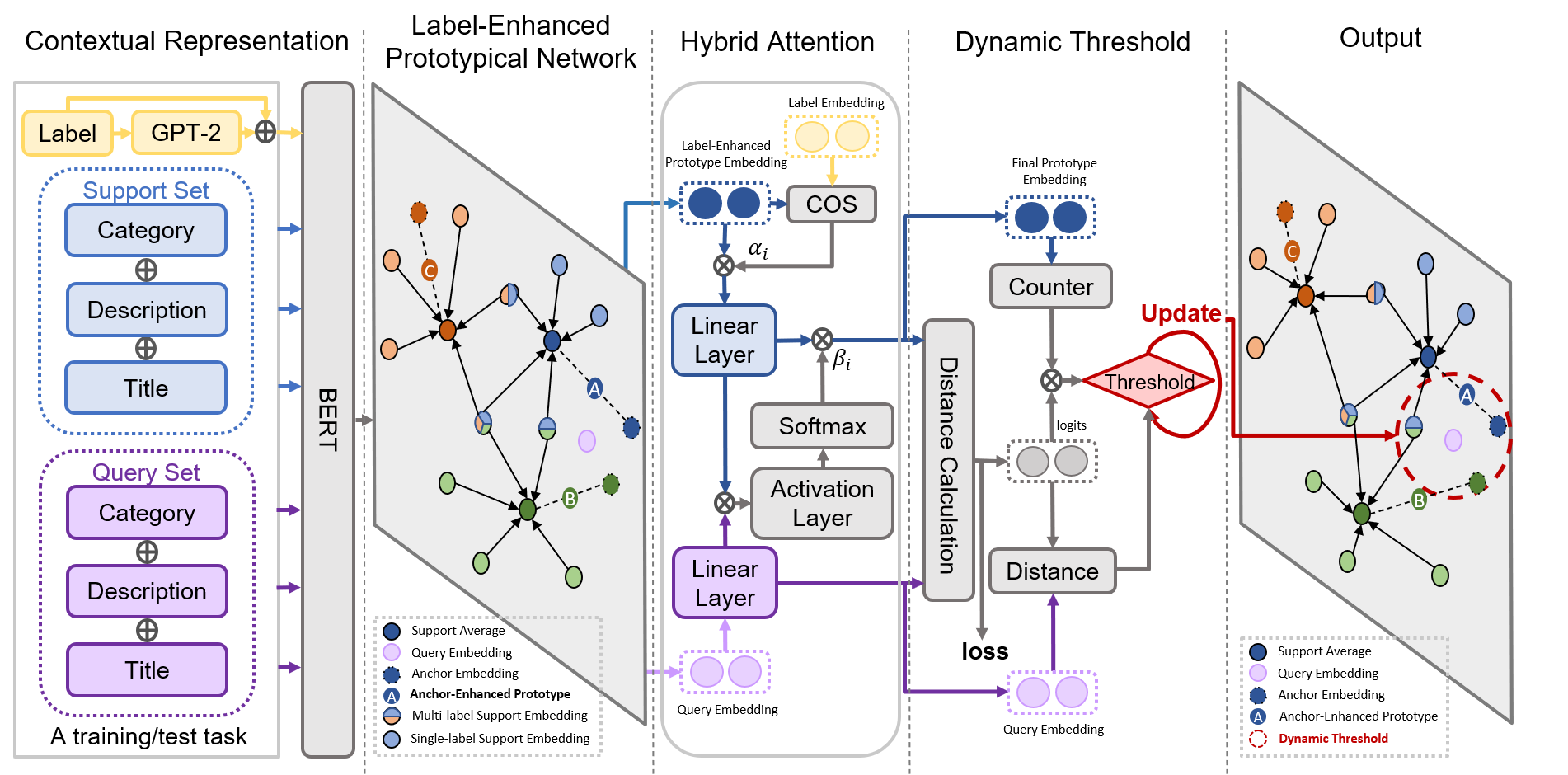}}
 \caption{\label{fig:framework} The overview of our proposed KEAF framework.}
 \vspace{-2mm}
 \end{figure*}

\vspace{-4mm} 
\subsection{Knowledge-Enhanced Attentive Framework}
In this section, we introduce the overview of KEAF in Figure~\ref{fig:framework}.

\subsubsection{Contextual Representations}~\label{sec:context}
Labels for AVE tasks are attribute value pairs such as `wallet type: long wallet', which may lose contextual information due to the simple format.
To achieve more information related to labels, we adopt GPT-2~\cite{radford2019language}
as the text generator to generate a detailed description for the attribute-value pairs. 
We adopt a pre-trained language model BERT~\cite{devlin-etal-2019-bert}
as the product input encoder to generate the contextual representation.
We construct a string [CLS;$c$;SEP;$t$;SEP;$d$] by concatenating product category, title and description as the input.
The output representation for the product input $r_{i} $ and label input $l_{i}$ is:
\begin{equation}~\label{equ:encoder}
r_{i} = tanh(W \cdot f_{\varnothing }(c_{i},t_{i},d_{i})+b), \quad
l_{i} = tanh(W \cdot f_{\varnothing }(g_{\varnothing}(l_{i}))+b)
\end{equation}
where $f_{\varnothing }$ is BERT encoder, $g_{\varnothing }$ is GPT-2 generator, $c$ is category, $t$ is title, $d$ is description and $l$ is `attribute is value' label information.

\subsubsection{Label-Enhanced Prototypical Network}
In Figure~\ref{fig:framework}, we adopt prototypical networks~\cite{snell2017prototypical} to get the original prototype of each attribute-value pair by averaging the embedding of support samples.
However, different labels may share the same support samples in multi-label settings, resulting in severe ambiguity.
To emphasize the difference between prototypes and reduce such ambiguity, we leverage label descriptions generated by GPT-2~\cite{radford2019language} to fully express the semantic information for attribute-value pairs and help learn more representative prototypes.
Label has shown significant effect on learning more discrimitative prototypes~\cite{hou2021few, liu2022label, zhao2022label}.
Thus, we combine the label with the average of support samples to compute a label-enhanced prototype $c_{i}$ with an interpolation factor $\eta$:
\begin{equation}~\label{equ:prototype}
c_{i}=\eta \times E(y_{i})+(1-\eta )\times \frac{1}{K_{i}}\sum_{j=1}^{K_{i}}E(x_{i}^{j})
\end{equation}
where $E(\cdot)$ is the BERT encoder, $y_{i}$ is the label description, $x_{i}^{j}\in \left \{ x|(x,Y)\in S\wedge y_{i}\in Y \right \}$ is the support sample labeled with $y_{i}$, and $K_{i}$ is shot number. The combination of label description and support embedding helps the prototypes better separated from each other.

\subsubsection{Hybrid Attention}
The aim of hybrid attention is to select more informative instances by retaining attribute-value relevant information while eliminating the negative effect triggered by the noise.
As shown in the third stage in Figure~\ref{fig:framework}, we first capture the similarity weight $\alpha_{i}$ in the label by calculating the semantic similarity between the label-enhanced prototype embedding $c_{i}$ from Equ~\ref{equ:prototype} and the attribute-value description embedding $l_{i}$ from Equ~\ref{equ:encoder}:
\begin{equation}
\alpha_{i} = cos(c_{i},l_{i}), \quad
\hat{c_{i}} = \alpha_{i} \times c_{i}
\end{equation}
where $cos(\cdot)$ is the cosine similarity and $\hat{c_{i}}$ gets the class-relevant information.
To further capture informative semantics from query-related instances and reduce the noise, we apply the instance-level attention, where each instance has a different importance factor $\beta_{i}$:
\begin{equation}\label{equ:logits}
\beta _{i}=\frac{exp(L(\hat{c_{i}})\times L(E(x_{i}^{q})))}{\sum_{{i}'=1}^{K}exp(L(\hat{c_{{i}'}})\times L(E(x_{{i}'}^{q})))}, \quad
\hat{r_{i}} = \beta_{i} \times \hat{c_{i}}
\end{equation}
where $L(\cdot)$ is the linear layer, $E(\cdot)$ is encoder from Equ.~\ref{equ:encoder}, $x_{i}^{q}$ represents the query instance and $\hat{r_{i}}$ is the final prototype.
Now, $\hat{r_{i}}$ contains label-relevant semantic information and it can be closer to the instances with features more related to queries.

\begin{table}[]
\small
\caption{Comparison with other multi-label FSL datasets.}
\vspace{-2mm}
\label{tab:dataset}
\centering
\begin{tabular}{lccc}
\hline
\multirow{2}{*}{Dataset} & \multirow{2}{*}{\begin{tabular}[c]{@{}c@{}}Train/Val/Test\\ \#Instance\end{tabular}} & \multirow{2}{*}{\begin{tabular}[c]{@{}c@{}}Train/Val/Test\\ \#Label\end{tabular}} & \multirow{2}{*}{\begin{tabular}[c]{@{}c@{}}Multi-label\\ Percentage\end{tabular}} \\
                         &                                                                                      &                                                                                   &                                                                                   \\ \hline
MS-COCO~\cite{10.1007/978-3-319-10602-1_48}                  & 97,600/-/24,400                                                                      & 64/-/16                                                                           & 38.81\%                                                                           \\
FewAsp~\cite{hu-etal-2021-multi-label}                   & 40,960/10,240/12,800                                                                 & 64/16/20                                                                          & 63.5\%                                                                            \\
TourSG~\cite{TourSG}                   & 19,351/1,600/4,800                                                                   & 68/17/17                                                                          & 18.13\%                                                                           \\
StanfordLU~\cite{eric-etal-2017-key}               & 3,517/2,512/2,009                                                                    & 14/10/8                                                                           & 16.57\%                                                                           \\
\textbf{MAVE}~\cite{10.1145/3488560.3498377}         & \textbf{29,458/-/2,049}                                                             & \textbf{45/-/17}                                                                  & \textbf{45.25\%}                                                                  \\
\textbf{Ours}         & \textbf{477,166/-/6,421}                                                             & \textbf{23/-/14}                                                                  & \textbf{43.38\%}                                                                  \\ \hline
\end{tabular}
\vspace{-4mm}
\end{table}

\begin{table*}[]
\small
\caption{Experimental results (\%) of multi-label few-shot learning on an in-house E-Commerce dataset.}
\vspace{-2mm}
\label{tab:result}
\centering
\begin{tabular}{l|cccccc|cccccc}
\hline
\multirow{3}{*}{Model} & \multicolumn{6}{c|}{1-shot}                                                                         & \multicolumn{6}{c}{5-shot}                                                                          \\  \cline{2-13} 
                       & Mac-P              & Mac-R              & Mac-F1             & Mic-P              & Mic-R              & Mic-F1             & Mac-P              & Mac-R              & Mac-F1             & Mic-P              & Mic-R              & Mic-F1             \\ \hline
Siamese~\cite{koch2015siamese}                & 12.52          & 30.84          & 16.07          & 12.11          & 30.23          & 17.18          & 22.16          & 25.76          & 21.75          & 21.38          & 24.55          & 22.76          \\
MTB~\cite{baldini-soares-etal-2019-matching}             & 13.00             & 36.86          & 17.70           & 12.77          & 36.76          & 18.87          & 10.16          & \textbf{98.89} & 18.06          & 10.12          & \textbf{98.51} & 18.35          \\
Proto\_BERT~\cite{snell2017prototypical}            & 24.71          & 39.73          & 28.00             & \textbf{30.44} & 41.82          & 34.78          & 30.71          & 40.39          & 32.81 & \textbf{32.85} & 42.55          & 36.86          \\
HCRP~\cite{han-etal-2021-exploring}            & 21.96          & 39.03          & 23.36          & 23.33          & 35.39          & 27.65          & 18.90           & 86.08          & 28.71          & 16.25          & 84.68          & 27.21          \\
FAEA~\cite{ijcai2022p407}            & 22.28          & 73.01          & 31.30           & 19.74          & 72.66          & 30.92          & 23.73          & 77.05          & 33.78          & 22.23          & 77.99          & 34.47          \\
SimpleFS~\cite{liu-etal-2022-simple}        & 14.75          & 69.33          & 23.12          & 18.08          & 72.80           & 28.93          & 16.81          & 62.65          & 25.16          & 21.59          & 66.69          & 32.55          \\ \hline
KEAF w/o att           & 24.79          & \textbf{73.57} & 34.43          & 24.54          & \textbf{75.35} & 36.92          & 26.39          & 73.50           & 36.69          & 26.01          & 75.08          & 38.48          \\
KEAF                   & \textbf{26.59} & 69.10           & \textbf{35.55} & 26.38          & 69.00             & \textbf{37.88} & \textbf{34.54} & 66.96          & \textbf{42.97} & 32.47          & 63.63          & \textbf{42.91} \\ \hline
\end{tabular}
\end{table*}

\begin{table*}[]
\small
\caption{Ablation result over components in 1-shot learning setting on in-house E-Commerce and MAVE datasets.}
\vspace{-2mm}
\label{tab:ablation}
\centering
\begin{tabular}{l|cccccc|cccccc}
\hline
\multirow{3}{*}{Model} & \multicolumn{6}{c|}{In-house E-Commerce}                                                                        & \multicolumn{6}{c}{MAVE}                                                                            \\  \cline{2-13} 
                       & Mac-P              & Mac-R              & Mac-F1             & Mic-P              & Mic-R              & Mic-F1             & Mac-P              & Mac-R              & Mac-F1             & Mic-P              & Mic-R              & Mic-F1           \\ \hline
w/o anchor weight      & 20.93          & 37.16          & 24.8           & 28.63          & 44.71          & 34.57          & 26.45          & 35.48          & 25.64          & 33.22          & 32.00             & 32.05          \\
w/o generator          & 24.59          & 47.75          & 29.32          & 27.59          & 49.27          & 35.25          & 28.56          & 32.70           & 26.77          & 40.29          & 33.98          & 36.08          \\
w/o threshold          & 24.64          & 44.66          & 29.37          & \textbf{28.68} & 48.26          & 35.57          &\textbf{32.99}            & \textbf{61.88}           & \textbf{38.22}            & 33.45            & \textbf{66.77}            & 44.27            \\
w/o category           & 23.76          & 56.62          & 30.98          & 27.34          & 60.32          & 37.38          & 18.82          & 23.83          & 17.88          & 35.19          & 21.66          & 26.27          \\
w/o attention          & 24.79          & \textbf{73.57} & 34.43          & 24.54          & \textbf{75.35} & 36.92          & 32.42 & 47.89          & 33.52          & 34.58          & 48.39          & 39.71          \\
KEAF (All)             & \textbf{26.59} & 69.10           & \textbf{35.55} & 26.38          & 69.00             & \textbf{37.88} & 31.38          & 50.92 & 34.47 & 37.65 & 55.54 & \textbf{44.46} \\ \hline
\end{tabular}
\end{table*}

\subsubsection{Dynamic Threshold}~\label{sec:dynamic}
In Figure~\ref{fig:framework}, we train the threshold $\tau$ by integrating the semantic information from both the support and query sets.
The thresholding function $T(\cdot)$ is calculated by the production of query label counter $\varphi(x_{i}^{q})$ with the relevance score between the final prototype $\hat{r_{i}}$ in Equ.~\ref{equ:logits} and query instance $r_{i}^{q}$ generated from Equ.~\ref{equ:encoder}.
The number of query labels is estimated by averaging the number of support labels of support instance $x_{i}$:
\begin{equation}~\label{equ:threshold}
\tau = T(\varphi(x_{i}^{q}), S) = \frac{1}{N\times K}\sum_{x_{i}\in X}^{}\varphi (x_{i})\odot  d(\hat{r_{i}},r_{i}^{q})
\end{equation}
where $S$ is the support set, $N$ and $K$ denotes N-way-K-shot, $\varphi(\cdot)$ represents the label counter, $\odot$ is element-wise production, and $d(\cdot)$ is the distance function.
The threshold is dynamically updated for each training epoch.
In testing, the framework predicts the query label set $Y_{i}^{q}$ by comparing 
the distance $d_{i}^{q}$ with the threshold $\tau$:
\begin{equation}
Y_{i}^{q} =\left \{ y_{i}^{q}|d_{i}^{q} < \tau , y_{i}^{q}\in Y \right \}
\end{equation}
The final threshold for the testing phase is chosen by the threshold value that has the best performance in the evaluation phase.
The model is trained by repeatedly sampling training episodes from $Y_{train}$ with support set $S$ and query set $Q$.
The model parameters are updated using the following binary cross entropy (BCE) loss:
\begin{equation}
\mathcal{L}=\sum_{I\in Q}^{}\sum_{i=1}^{N}y_{i}^{I}\cdot log\sigma (q_{i}^{I})+(1-y_{i}^{I})\cdot log(1-\sigma (q_{i}^{I}))
\end{equation}
where $Q$ is query shot, $N$ is N-way, $\sigma(\cdot)$ is the sigmoid function and $y_{i}^{I}$ represents the ground truth.

\section{Experiments}
\subsection{Experimental Setup}~\label{sec:data}
We evaluate our model over two datasets: a large e-Commerce platform in Japan
and MAVE~\cite{10.1145/3488560.3498377}.
The dataset statistics is shown in Table~\ref{tab:dataset}. 
We compare KEAF with SOTA few-shot IE approaches, models without label information: Siamese~\cite{koch2015siamese}, Proto~\cite{snell2017prototypical}, MTB~\cite{baldini-soares-etal-2019-matching}, and models with label information: HCPR~\cite{han-etal-2021-exploring}, FAEA~\cite{ijcai2022p407} and SimpleFSRE~\cite{liu-etal-2022-simple}.
For evaluation, we use both Micro(Mic-) and Macro(Mac-) Precision, Recall and F1.
The max length of input is 512 for project and 32 for anchor.
The dimension size is 768.
We vary the label interpolation factor $\eta$ in $\begin{Bmatrix}0.1, 0.5, 1.0\end{Bmatrix}$ and the optimal anchor weight is selected.
Our model is implemented on PyTorch and optimized with AdamW optimizer.
The learning rate is $10^{-5}$ with weight decay $10^{-6}$.
The batch size is 1 and dropout rate is 0.2.
The experiments are conducted on Nvidia A100 GPU with 80G GPU memory.

\subsection{Results and Discussions}
\subsubsection{Main Results}
\begin{table}[]
\small
\caption{Results of F1 score (\%) on MAVE dataset.} 
\vspace{-2mm}
\label{tab:mave}
\centering
\begin{tabular}{l|cc|cc}
\hline
Model           & \multicolumn{2}{c|}{1-shot}     & \multicolumn{2}{c}{5-shot}                        \\ \cline{2-5} 
                & Macro-F1       & Micro-F1       & Macro-F1               & Micro-F1                \\ \hline
Siamese~\cite{koch2015siamese}         & 15.83          & 18.55          & 28.29                   & 28.15                   \\
MTB~\cite{baldini-soares-etal-2019-matching}      & 17.36          & 18.26          & 20.62                   & 20.76                   \\
Proto\_BERT~\cite{snell2017prototypical}     & 30.09          & 36.79          & 33.56                   & 39.46                   \\
HCRP~\cite{han-etal-2021-exploring}     & 18.55          & 19.31          & 16.11                   & 16.58                   \\
FAEA~\cite{ijcai2022p407}     & 16.38          & 16.87          & 16.63                   & 17.11                   \\
SimpleFS~\cite{liu-etal-2022-simple} & 26.63          & 30.43          & 17.44                   & 23.05                   \\ \hline
KEAF w/o att    & 33.52          & 39.71          & \textbf{38.22}          & 44.27          \\
KEAF            & \textbf{34.47} & \textbf{44.46} & {36.40} &\textbf{44.33} \\ \hline
\end{tabular}
\vspace{-6mm}
\end{table}

The results of multi-label FSL are shown in Table~\ref{tab:result} and Table~\ref{tab:mave}.
We observe: (1) KEAF significantly outperforms other baselines on both macro and micro F1 in 1-shot and 5-shot settings.
These results reveal that KEAF better learns the prototypes and captures the informative semantics.
(2) On the in-house E-Commerce dataset, models using label semantics improve the performance more in 1-shot than 5-shot setting.
This is consistent with our expectations that adding label information helps reduce ambiguity.
On MAVE, baselines using labels even have a worse performance.
We conjecture that the original labels in MAVE are too simple to learn anything, and they even cause noise.
In KEAF, we generate a more detailed label description for better integrating the label and reducing the noise, resulting in the best performance among all models.
(3) MTB and other baselines demonstrate good results only on Recall and bad results on Precision.
For AVE, low precision means lots of human efforts are needed to manually remove the extracted non-relevant attribute-value pairs.
A possible reason is that a very large threshold is learned on these baselines, trying to predict as many labels as possible and resulting in a very large recall value. 
In contrast, KEAF better learns the threshold and balance the Precision and Recall, resulting in the highest F1 score.

\subsubsection{Ablation Study}
To verify the effectiveness of each component in KEAF, we conduct the 1-shot ablation study in Table~\ref{tab:ablation}.
We observe: (1) Fusing anchor to the prototypes results in a large performance improvement because the label helps discriminate prototypes and reduce ambiguity. 
(2) Generating a more detailed label description helps improve the performance more on MAVE than on the in-house E-Commerce dataset.
We conjecture that MAVE is an English dataset and English GPT-2 is well trained than Japanese GPT-2, resulting in a more accurate label description on MAVE. 
(3) Category information shows vital importance on MAVE.
We think that AV pairs are from different categories and adding the category can better separate the prototypes.
(4) Using the attention can improve the performance by reducing the noise to some extent.

\section{Conclusion}
In this paper, we formulate attribute value extraction task in few-shot learning to solve the long-tailed data problem and limited training data for new products.
We propose a Knowledge-Enhanced Attentive Framework for product AVE.
We design a label-enhanced prototypical network with the hybrid attention to alleviate ambiguity and noise, and capture more informative semantics.
We train a dynamic threshold to achieve multi-label inference.
Results demonstrate that KEAF outperforms other SOTA IE models significantly. 

\bibliographystyle{ACM-Reference-Format}
\balance
\bibliography{sample-base}

\appendix

\end{document}